\documentclass[aps,prl,preprint,groupedaddress]{revtex4-1}

\usepackage{graphicx}
\usepackage{dcolumn}
\usepackage{bm}
\usepackage{mathtools}

\begin{document}


\title{Construction of controlled-NOT gate based on microwave-activated phase (MAP) gate
in two transmon system}


\author{Taewan Noh$^{1}$, Gwanyeol Park$^{1,2}$, Soon-Gul Lee$^{2}$, Woon Song$^{1}$, Yonuk Chong$^{1,3}$}
\affiliation{$^{1}$Korea Research Institute of Standards and Science, Daejeon 34113, Republic of Korea \\
$^{2}$Korea Universitiy Sejong campus, Sejong 30019, Republic of Korea\\
$^{3}$University of Science and Technology, Daejeon 34113, Republic of Korea}


\date{\today}

\begin{abstract}
We experimentally constructed an all-microwave scheme for the controlled-NOT (cNOT) gate between two superconducting transmon qubits in a three dimensional cavity. Our cNOT gate is based on the  microwave-activated phase (MAP) gate, which requires
an additional procedure to compensate the accumulated phases during the operation of the MAP gate. We applied Z-axis phase gates using microwave hyperbolic secant pulse on both qubits with adequate rotation angles systematically calibrated by separate measurements.
We evaluated the gate performance of the constructed cNOT gate by performing two-qubit quantum process tomography (QPT). Finally,
we present the experimental implementation of the Deutsch-Jozsa algorithm using the cNOT gate.
\end{abstract}

\pacs{}

\maketitle


\section*{Introduction}

During the last decades, superconducting qubits coupled to a microwave cavity has been the key building block for realizing a scalable quantum processor. Remarkable improvements in qubit coherence, fidelity of qubit operation and microwave control \cite{barends, sheldon1, chen, sheldon2, corcoles, nori1}, have enabled many demonstrations of quantum algorithms \cite{dic1, yama, mariantoni} in superconducting qubit systems.
In order to implement quantum algorithms, a universal quantum gate set is required. Therefore, it is essential to realize controlled entagling gates between two qubits with high fidelity and efficient operation. Various schemes for entangling gates have been demonstrated, which includes the higher-level resonance induced dynamical c-Phase gate \cite{dic1, yama, strau} and the cross-resonance gate \cite{corc, parao, rigetti, chow3}. While the three dimensional (3D) transmon system \cite{paik} has advantage in showing a long coherence time, flexibility in qubit-resonator configuration, and relatively simple fabrication requirements, it is more suitable to fixed-frequency qubits, and only a few entangling schemes \cite{paik2, chow1, poletto} for fixed-frequency qubits have been experimentally demonstrated. Meanwhile, all-microwave scheme-based gates can be operated with a single high-frequency cable without additional tuning wirings, so that they are more hardware-efficient.  Therefore, by studying the microwave-activated phase (MAP) gate {\cite{chow1}}in more detail and constructing the controlled-NOT (cNOT) gate in a more efficient way, we hope that this study will be helpful for a realizing small-scale quantum processor in circuit QED system based on three dimensional microwave cavity.

In this paper, we present an all-microwave scheme for a cNOT gate between two transmons embedded in a single three dimensional microwave cavity. The cNOT gate
consists of the MAP gate followed by the Z-axis phase gate on each qubit with systematically calibrated phases. This combination for the realization of the cNOT has advantage over the previous refocusing scheme \cite{paik2} in that the total gate time can be reduced and the coherence-time limited gate infidelity can be mitigated. 
By performing two-qubit quantum process tomography (QPT), we obtained the process fidelity of 0.65 for our cNOT gate. Then we applied the cNOT gate to demonstrate the two-qubit version of the Deutsch-Josza algorithm.

\section*{Experimental results}

\subsection*{Realization of the MAP gate}

The MAP gate demonstrated by J. Chow \textit{et 
al.}\cite{chow1} takes advantage of the always-on interaction between two non-computational states. When the energy levels of those states are aligned, the interaction is given via a common cavity mode capacitively coupled to each qubit. Because the gate speed is proportional to the coupling, the alignment of two states is important in order to maximize their coupling.
In our experiment, instead of using two pre-selected fixed-frequency qubits as in ref.\cite{chow1}, we used a tunable-frequency qubit for one of the qubits so that we are able to tune the qubit frequency {\textit{in-situ}} to an optimal value. In order to apply magnetic flux for frequency tuning, we used a copper cavity, in which two (one tunable and one fixed) transmon qubits fabricated on separate silicon substrates were mounted, as shown in the top left panel of Fig. 1(a).
If we label the tunable-frequency qubit as Q1 and
the fixed-frequency qubit as Q2 for the rest of this paper, $|12\rangle$ and $|03\rangle$ states (where the first (second) number in the bracket stands for the state of Q1 (Q2)) are two specific states that need to be aligned in the MAP gate scheme. By performing the three-tone microwave spectroscopy, we carefully tune the magnetic flux bias through the tunable qubit (Q1) to align these two states, as plotted in Fig.1(a). Two continuous wave (CW) microwave tones with the frequencies respectively corresponding to $\omega_{01}$ and $\omega_{12}$ of Q2 populates $|02\rangle$ state, while the third tone is swept to observe the transitions from $|02\rangle$ to $|12\rangle$ and $|03\rangle$ states by monitoring the $|02\rangle$ state population. At the magnetic flux value optimally aligning $|12\rangle$ and $|03\rangle$ states, the resonant frequencies $\omega_{01}$ of each qubit are $\omega_{01} (\rm{Q_1})$ = 5.6498 GHz and $\omega_{01} (\rm{Q_2})$ = 6.2903 GHz, respectively. In addition, the resonant frequencies between the first and the second excited states $\omega_{12}$ are $\omega_{12} (\rm{Q_1})$ = 5.3336 GHz and $\omega_{12} (\rm{Q_2})$ = 5.9852 GHz, respectively. Therefore the anharmonicities of each qubit are $\alpha(\rm{Q_1})$ = -316 MHz and $\alpha(\rm{Q_2})$ = -305 MHz. The coherence times are measured to be $T_1$ = 21 $\mu$s, $T_2^{*}$ = 5 $\mu$s for Q1 and $T_1$ = 15 $\mu$s, $T_2^{*}$ = 11 $\mu$s for Q2. Since the resonant frequencies of both qubits are far detuned from that of the cavity, $\omega_r$= 7.16207 GHz, and the detuning is larger compared to the coupling $g$ (which was estimated to be 122 MHz from a separate measurement), our system lies in the dispersive-coupling  regime. We used the low-power dispersive readout to monitor the qubit states.
Here we briefly explain the MAP gate. As depicted
in Fig.1(a), when $|12\rangle$ and $|03\rangle$ states are aligned, the splitting between these states occurs due to the interaction of two qubits. The magnitude $\delta$ of the splitting is measured to be 15 MHz for the present experiment, as shown in Fig.1(a). We denote the frequency difference between $|01\rangle$ and $|02\rangle$ states as $\epsilon$ (which is basically the  frequency difference $\omega_{12}$ between the first excited state $|1\rangle$ and the second excited state $|2\rangle$ of Q2), while the frequency difference between $|11\rangle$ and $|12\rangle$ states as $\epsilon'=\epsilon-\delta/2$. If we apply a microwave tone (Stark tone) with a frequency close to $\epsilon$ (thus close to $\epsilon'$ at the same time), the ac-Stark shifts of the energy levels occur to the states including two of the computational states, $|01\rangle$ and $|11\rangle$. For the data presented in this paper, we set the frequency of the Stark tone as 6.0152 GHz, which is detuned from $\omega_{12}(\rm{Q_2})$ by 30 MHz. It should be noted that the amount of the ac Stark shifts $\delta \epsilon$ for
$|01\rangle$ state and $\delta \epsilon'$ for $|11\rangle$ state are different, \textit{i.e.},
$\delta \epsilon \neq \delta \epsilon'$, since the detuning of the Stark tone is different due to the existence of the splitting $\delta$. This
is the basic principle of the MAP gate. The amount of the energy shift induced by the Stark tone can be observed by a Ramsey experiment. With the state of Q1 at $|0\rangle$, the Ramsey experiment on Q2 exhibits the oscillation of the population between the states $|00\rangle$ and $|01\rangle$ with the oscillation frequency given by $\delta \epsilon$. On the other hand, with the state of Q1
in $|1\rangle$, the oscillation occurs between the states $|10\rangle$ and
$|11\rangle$ but now the oscillation frequency is $\delta \epsilon'$. Due to the difference between $\delta \epsilon$ and $\delta \epsilon'$, two traces of the Ramsey experiment would be completely out-of-phase at a certain time $t_g$. As we increase the power of the Stark tone  or make the frequency of the Stark tone closer to $\omega_{12}$ of Q2, the Ramsey oscillation becomes faster and $t_g$ becomes shorter. Short $t_g$ is advantageous considering the finite coherence time of the qubits. However, the leakage to higher non-computational states due to the strong Stark drive tone develops at the same time, hence there exists a trade-off between the time $t_g$ and the infidelity caused by the leakage.

\subsection*{Implementation of Z-axis phase gate based on hyperbolic secant pulse}
Applying the MAP gate for $t_g$ results in a conditional operation depending on the state of Q1, but at the same time, a certain amount of the additional phase accumulation occurs during the operation which depends on the initial state. For example, the four initial states $|00\rangle$, $|01\rangle$, $|10\rangle$, and $|11\rangle$ are transformed to $e^{i\phi_{00}}|00\rangle$, $e^{i\phi_{01}}|01\rangle$, $e^{i\phi_{10}}|11\rangle$, and $e^{i\phi_{11}}|10\rangle$, respectively. We only need to take into account the relative phase accumulation by setting the global phase $\phi_{00}=0$, and then by rewriting phases as $\phi=\phi_{01}$, $\phi'=\phi_{10}$, and $\phi+\phi'=\phi_{11}$, the matrix representation of the MAP gate can be written as

\begin{equation}
U_{{\rm{MAP}}}=
\left( \begin{array}{cccc}
1 & 0 & 0 & 0 \\
0 & e^{i \phi} & 0 & 0 \\
0 & 0 & 0 & e^{i (\phi+\phi')} \\
0 & 0 & e^{i \phi'} & 0 \\
\end{array}
\right)
\end{equation}
  
in {$|00\rangle$, $|01\rangle$, $|10\rangle$, and $|11\rangle$} basis. In order to construct the cNOT gate out of the MAP gate, the phases $\phi$ and $\phi'$ need to be compensated. The compensation of the phases has been realized in the experiments by the current pulse-based Z-gate\cite{dic1, yama} or the refocusing method \cite{paik2}. While the current pulse method is only applicable to dynamically-tunable qubit systems, the refocusing method can be applied to fixed-frequency qubits. Although our system includes a tunable-frequency qubit, our qubits are operated as fixed-frequency qubits once we fix the flux bias at its optimal value.  Since the refocusing method requires application of at least two conditional gates, there is a limit in gate fidelity improvement due to the long operation time, especially when the coherence time is not long enough (which is usually the case). This would be more prominent as the number of qubits in the system increases. Therefore, we adopt a different scheme for compensating unwanted phases. The Z-axis phase gate recently reported by H. Ku \textit{et al.}\cite{ku} takes advantage of the unique property of the hyperbolic secant pulse.\cite{economou} Unlike other frequently used pulse shapes such as square or Gaussian, hyperbolic secant shape envelop on a microwave tone with frequency $\omega_D$ can drive Rabi oscillations of a two-level system with the oscillation period independent of the detuning of microwave frequency  $\Delta = \omega_D - \omega_{01}$ as shown in Fig. 2(a). Therefore, regardless of the drive frequency $\omega_D$, an arbitrary initial state
$|\Psi\rangle = a|0\rangle + b|1\rangle$ should maintain the probability of being at the ground state (excited state) as $|a|^2$ ($|b|^2$) after a $2 \pi$ cyclic evolution driven by hyperbolic secant pulse. This leads the wavefunction to $a|0\rangle + e^{i\phi}b|1\rangle$ with $\phi$ being determined by the detuning $\Delta$. This is the
principle of the Z-axis phase gate with a single parameter $\Delta$.

Experimental demonstration and calibration of the Z-axis phase gate was conducted as illustrated in the inset of the Fig. 2(b) : i) Rotate the initial state $|0\rangle$ by $\pi/2$ along Y axis. ii) Apply a hyperbolic secant pulse corresponding to $2 \pi$ rotation. iii) Perform single-qubit quantum state tomography (QST) to estimate the expectation value of $\langle X\rangle$, $\langle Y\rangle$, and $\langle Z\rangle$ of the resulting quantum state on the Bloch sphere. iv) Repeat from i) to iii)
by varying the frequency of the microwave $\omega_D$ applied with the hyperbolic secant pulse in ii). The plot in Fig. 2(b) exhibits the result of the Z-axis phase gate implemented on Q1 of our system for three different pulse lengths 200 ns, 300 ns, and 400 ns ($2\pi$ cyclic rotation). The envelope of hyperbolic secant pulse extends over $\pm 4\sigma$ where $\sigma$ is the standard deviation. For all cases, while the polar angle $\theta$ with respect to Z-axis remains constant at $\pi/2$, azimuthal angle $\phi$ is rotated around Z-axis as a function of the detuning $\Delta$. The dependence of $\phi$ on the detuning $\Delta$ shows good agreement with the expected relation, $\phi = 4 {\rm{arctan}}(\Delta/\rho)$, where the bandwidth $\rho$ is related to the standard deviation of the hyperbolic secant pulse $\sigma$ by $\sigma = \pi/(2\rho)$.\cite{ku} The fidelity of the quantum state at each value of the detuning can be extracted by using the formula $F = {\rm{Tr}}[\rho_{the}\rho_{exp}]$, where $\rho_{the}$ and $\rho_{exp}$ are the theoretical and experimental density matrix of the target quantum state respectively. The estimated fidelity is shown in Fig. 2(c) for the range of the detuning -5MHz $\leq \Delta \leq$ 5MHz for both Q1 and Q2, which exhibits the average of the state fidelity roughly 0.95 for Q1 and 0.97 for Q2.  

Now we implement the Z-axis phase gate in our system to compensate the phases accumulated during the MAP gate. The combination of the pulse length of Z-axis phase gate and the corresponding detuning for a specific rotation angle can be chosen arbitrarily. However, according to Ku \textit{et al}.,\cite{ku} large detuning can cause infidelity. Thus, we have chosen 400 ns as the pulse length with which wan can produce larger angle rotation with smaller detuning, while the fidelity remains the same as the shorter pulse length cases. We then attempt first to compensate the phase $\phi$ in the matrix $U_{\rm{MAP}}$ shown above, which can be observed by applying the MAP gate on an initial state $\frac{1}{\sqrt{2}} (|00\rangle+|01\rangle)$, resulting in $\frac{1}{\sqrt{2}} (|00\rangle+e^{i \phi}|01\rangle)$. Adding
Z-axis phase gate on Q2 (Z$_2$ gate), after the MAP gate permits us to control $\phi$ with the detuning $\Delta_2$ of the microwave frequency for the Z$_{2}$ gate. Then the final state can be estimated by performing two-qubit QST at each value of $\Delta_2$. By this procedure we are able to determine the value of $\Delta_2$ for compensating $\phi$.
However, the large number of measurements required to run the entire two-qubit QST protocol is a big overhead. Instead, we used only a few measurements which are sensitive to the modulation of $\phi$. For example, as shown in the top left plot in Fig. 3(a), among 36 over-complete set of pre-pulses $\{I, X_{\pi}, X_{\pm \pi/2}, Y_{\pm \pi/2}\} \otimes \{I, X_{\pi}, X_{\pm \pi/2}, Y_{\pm \pi/2}\}$ prior to the readout of two-qubit QST, the one with $I \otimes I$ pre-pulse is insensitive to the evolution of the state $\frac{1}{\sqrt{2}} (|00\rangle+e^{i \phi}|01\rangle)$ when the phase $\phi$ is modulated (black open circle). In contrast, the one with different pre-pulses such as $X_{\pi/2}\otimes X_{\pi/2}$ (orange open triangle) or $Y_{\pi/2}\otimes Y_{\pi/2}$ (navy open square) is sensitive. The theoretical estimation is based on the actual calibration of the joint two-qubit readout measurement operator \cite{filipp,chow4}, 
${\rm{M}} = \beta_{II} II+ \beta_{ZI} ZI+ \beta_{IZ} IZ +\beta_{ZZ} ZZ$ with  $\beta_{II}$= 2.72, $\beta_{ZI}$= 1.1,  $\beta_{IZ}$= 0.86,  and $\beta_{ZZ}$= 0.44 in our specific measurement setup. While we measure the population of the state being at $|00\rangle$, there is a finite contribution to the readout from other computational states, $|01\rangle$, $|10\rangle$, and $|11\rangle$, which are represented as the $\beta$ parameters and are calibrated from the experiment for the configuration of our readout pulse.  For example, in case of applying  $I \otimes I$, $X_{\pi/2}\otimes X_{\pi/2}$ and $Y_{\pi/2}\otimes Y_{\pi/2}$ pre-pulses, 
the readouts of the final state are represented as $\beta_{II}+\beta_{ZI}$, $\beta_{II} + \beta_{IZ} \cdot \rm{sin} \phi$ and $\beta_{II} + \beta_{IZ} \cdot \rm{cos} \phi$, respectively. Therefore, the readout of the final state exhibits the minimum value at $\phi = 0 $ and the maximum at $\phi = \pm \pi$ when $Y_{\pi/2}\otimes Y_{\pi/2}$ pre-pulse is applied, which corresponds to $\frac{1}{\sqrt{2}} (|00\rangle+|01\rangle)$ and $\frac{1}{\sqrt{2}} (|00\rangle-|01\rangle)$, as shown in the left plot in Fig. 3(a). In the experiment, we measured the evolution of the state $ \frac{1}{\sqrt{2}} (|00\rangle+e^{i\phi}|01\rangle)$ as we modulate the phase $\phi$ with the gate Z$_2(\Delta_2)$ by applying two of the QST protocols including the pre-pulses $X_{\pi/2}\otimes X_{\pi/2}$ and $Y_{\pi/2}\otimes Y_{\pi/2}$, as shown in the right plot in Fig. 3(a). The dotted lines are the theoretical estimations borrowed from the left plot with the conversion between the phase $\phi$ and the detuning $\Delta_2$ obtained from a separate measurement in Fig. 2(b). Qualitative agreement of the measured data with the theoretical estimation is observed in the plot. The discrepancy arises mainly due to the decoherence caused by the long operation time of the MAP gate, compared to the coherence times of both Q1 and Q2, before applying any set of prepulses. Then we performed the entire protocols of two-qubit QST for specific values of $\Delta_2$ = -0.2 MHz (near 0 in terms of the rotation angle induced by the Z$_2$) and 3.0 MHz (near $\pi$). We confirmed that the density matrix of the state indeed appeared as the one for $\frac{1}{\sqrt{2}} (|00\rangle+|01\rangle)$ and $\frac{1}{\sqrt{2}} (|00\rangle-|01\rangle)$ respectively, as shown in the right side of Fig. 3(a). The state fidelity of the resulting quantum state is 0.75 for both detunings according to the estimation by QST. It should be noted that the resulting angle $\phi$ after applying the MAP gate on the state $\frac{1}{\sqrt{2}}(|00\rangle + |01\rangle)$ is very close to zero only for this specific choice of the power of the Stark tone, but this should not be always the case.  We determine the optimal value of $\Delta_2$ for compensating the phase $\phi$ as -0.2 MHz for the current setup of the experiment.  
 
The phase $\phi'$ can be compensated in a similar manner. By preparing an initial state $\frac{1}{\sqrt{2}} (|00\rangle+|10\rangle)$ and applying the MAP gate, we generate $\frac{1}{\sqrt{2}} (|00\rangle+e^{i \phi'}|11\rangle)$.
In addition to the Z$_2$ gate with the detuning fixed at the optimal value $\Delta_2$ = -0.2 MHz, the Z-axis phase gate on Q1, namely Z$_1$ gate, is applied after the MAP gate together with the Z$_2$ gate. Because our goal is to compensate the total accumulated phase in the end, this sequence of compensating the phase of each qubit one by one on top of other qubit’s compensation guarantees that {\textit{n}}-measurements will complete the calibration for {\textit{n}}-qubit case in general. In order to find the optimal value for $\Delta_1$, we perform two-qubit QST with the pre-pulses $X_{\pi/2}\otimes X_{\pi/2}$ and $Y_{\pi/2}\otimes Y_{\pi/2}$ as we modulate the phase $\phi'$ with the Z$_1 (\Delta_1)$ gate. Theoretically, in case of 
applying  $I \otimes I$, $X_{\pi/2}\otimes X_{\pi/2}$ and $Y_{\pi/2}\otimes Y_{\pi/2}$ pre-pulses, 
the readouts of the final state are represented as $\beta_{II}+\beta_{ZZ}$, $\beta_{II} - \beta_{ZZ} \cdot \rm{cos} \phi'$ and $\beta_{II} + \beta_{ZZ} \cdot \rm{cos} \phi'$, respectively. Therefore, the readout of the final state exhibits the minimum value at $\phi' = 0 $ and the maximum at $\phi' = \pm \pi$ when $X_{\pi/2}\otimes X_{\pi/2}$ pre-pulse is applied, which corresponds to $\frac{1}{\sqrt{2}} (|00\rangle+|11\rangle)$ and $\frac{1}{\sqrt{2}} (|00\rangle-|11\rangle)$. On the other hand, the maximum value appears at $\phi' = 0 $ and minimum appears at $\phi = \pm \pi$ when $Y_{\pi/2}\otimes Y_{\pi/2}$ pre-pulse is applied, which is shown in the left plot in Fig. 3(b). The right plot shows the experimental results which exhibits a qualitative agreement with the theoretical expectation. With $\Delta_1$ fixed at 2.8 MHz (0.85$\pi$ in terms of the rotation angle) and -0.8 MHz (-0.15$\pi$) respectively, which correspond to the maximum and the minimum of the trace obtained in case of $Y_{\pi/2}\otimes Y_{\pi/2}$ pre-pulses, we perform the entire protocol of two-qubit QST. The reconstructed density matrix represents $\frac{1}{\sqrt{2}} (|00\rangle+|11\rangle)$ and $\frac{1}{\sqrt{2}} (|00\rangle-|11\rangle)$ as expected. Here the state fidelities are estimated to be 0.75 from the QST result. This leads us to determine the optimal value of $\Delta_1$ as 2.8 MHz for compensating the phase $\phi'$. As the final check, we apply the MAP gate to an initial state $\frac{1}{\sqrt{2}} (|01\rangle+|11\rangle)$ and observe that the final state results in $\frac{1}{\sqrt{2}} (|01\rangle+|10\rangle)$ by the addition of the Z$_1$ and Z$_2$ gates with the optimal detuning values $\Delta_1$ = 2.8 MHZ and $\Delta_2$ = -0.2 MHz. This confirms the cancellation of the phase factor $e^{i (\phi + \phi')}$ in the matrix $U_{{\rm{MAP}}}$.

\subsection*{Evaluation of the cNOT gate and demonstration of the Deutsch-Jozsa algorithm}
Now we will confirm that we constructed the cNOT gate based on the MAP gate, and estimate the process fidelity of the Z-gate and the cNOT gate. Quantum process tomography
(QPT) is a protocol for such purpose, which is used to analyze any arbitrary gate in the system. We adopted the protocol used in Ref.\cite{chow2} where
the Pauli transfer matrix $\textit{\textbf{R}}$ mapping the input
Pauli state vector $\vec{p}_{in}$ in the basis of two-qubit Pauli operator into the output Pauli state vector $\vec{p}_{out}$, \textit{i.e.}, $\vec{p}_{out} = \textit{\textbf{R}} \vec{p}_{in}$, is extracted. QPT consists of preparation of initial state, application of the gate of interest, and QST in order to tomographically reconstruct the final state by 36 over-complete set of pre-pulses prior to the readout pulse. Since we utilize the same 36 pre-pulses to prepare the initial states,
QPT for a single gate of interest is accomplished by
running 36$\times$36 = 1296 measurements. Then the experimental Pauli
transfer matrix $\textit{\textbf{R}}$ is estimated via maximum-likelihood
estimation in which $\textit{\textbf{R}}$ should take the form of a physical matrix that satisfies the conditions of trace preserving and complete positivity. 

We implemented the protocols for two-qubit QPT in our system and tested several gates including single-qubit gates and two-qubit gates. For example, in addition to the rotation gate along X and Y axes, we performed QPT for the Z$_1 (\Delta_1)$ and the Z$_2 (\Delta_2)$ gate. These gates play crucial roles in constructing the cNOT gate in our experiment, therefore, the fidelities of those Z-phase gates are to be estimated. Fig. 4(a) shows the Pauli transfer matrix $\textit{\textbf{R}}$ extracted for the special cases of $\Delta_1$ and $\Delta_2$ corresponding to the gate $Z_{\pi/2} \otimes I$ (left) and $I \otimes Z_{\pi/2}$ (right). The ideal matrices $\textit{\textbf{R}}$ are shown together for comparison. For this example, we chose the pulse length of both Z$_1$ and Z$_2$ gates as 300 ns with the detuning $\Delta_1$ = 1.85 MHz for Z$_1$ gate and $\Delta_2$ = 1.65 MHz for Z$_2$ gate (rotation angle = $\pi/2$). The process fidelity defined as $F_g = {(\rm{Tr}}[R^{\dagger}_{\rm{ideal}}R_{\rm{exp}}]+d)/(d^2+d) $ \cite{chow2} with $d = 4$ for two-qubit system is estimated to be 0.90 for both $Z_{\pi/2} \otimes I$ and $I \otimes Z_{\pi/2}$. Based on the experimental results of the state fidelity shown in Fig. 2(c), we expect the process fidelity to be more or less the same for different pulse lengths 200 ns, 300 ns, and 400 ns. It should be noted that the fidelity estimation by QPT gives lower value compared to QST results. We think that the difference of QPT and QST estimation of fidelity  can be attributed to possible drift in the system or intermittent noise during the long pulse sequence of the entire protocol of QPT, which includes over thousand measurements. To be more specific, running the protocol of QST usually takes less than 10 minutes in our system whereas it takes more than 4 hours to run the entire protocol of QPT. 
For the two-qubit gate, the schematic pulse sequence of two-qubit QPT when the gate of interest is the cNOT gate is illustrated in the top panel of Fig. 4(b). As we explained above, the cNOT gate consists of the MAP gate with the gate time $t_g$ of 1070 ns followed by the Z-phase gate of 400ns gate time on each qubit with $\Delta_1$ = 2.8 MHz and $\Delta_2$ = -0.2 MHz for the phase compensation. The bottom panel of Fig. 4(b)reveals the Pauli transfer matrix $\textit{\textbf{R}}$ reconstructed from the experimental data. By comparing the experimental $\textit{\textbf{R}}$ with the $\textit{\textbf{R}}$ shown on the right side, we obtained the process fidelity of 0.65 for our cNOT gate by QPT. 
As a further demonstration of the cNOT gate we constructed, we applied it to the Deutsch-Jozsa algorithm. \cite{dj} In the simplest form of the Deutsch–Jozsa algorithm, one can determine whether an unknown function is a constant function or a balanced function by doing a single query. We implemented two-qubit version of the algorithm, where there are two constant functions ($f_0(x)=0$ and $f_1(x)=1$) and two balanced functions ($f_2(x)=x$ and $f_3(x)=1-x$).
Fig. 5(a) describes the gate sequence for running the Deutsch-Jozsa algorithm. The encoding unitaries $U_i$ for four functions are $U_0 = I \otimes I$, $U_1 = I \otimes X_{\pi}$
$U_2 =\rm{cNOT}$, and $U_3 = (Y_{\pi} \otimes I){\rm{cNOT}}(Y_{-\pi} \otimes I)$. The results of the Deutsch-Jozsa algorithm are summarized in Fig. 5(b) as the density matrices of the output states. Although the contrast for the balanced functions $f_2$ and $f_3$ are lower than those of the constant functions $f_0$ and $f_1$ due to the infidelity of the cNOT gate, the result is consistent with the theoretical prediction. This again confirms correct function of the constructed cNOT gate with the combination of the Z-axis phase gates and the MAP gate.  

\section*{Discussion} 
  
In summary, we constructed the cNOT gate between two fixed-frequency transmon qubits embedded in a three dimensional microwave cavity. The cNOT gate is based on the MAP gate which is an all-microwave entangling gate scheme. We used \textit{in-situ} tuning of one of two qubits in order to optimally align the energy levels between two transmons. The tunability is a requirement for the optimal MAP gate operation and also enables detailed study of the MAP gate. Then the cNOT gate is realized by adding the sech-pulse based Z-axis phase gate on each qubit with an appropriate rotation angle. This compensates the accumulated phase on the computational states during the operation of the MAP gate. We adopted a protocol for the calibration of the accumulated phases from the MAP gate which works for any entangling c-phase gate in general. This method of adding Z-axis phase gate provides a better phase compensation scheme compared to the refocusing method used in previous studies, in that this scheme can benefit from reduction of the number of pulses and the total gate time. We performed two-qubit QPT on our cNOT gate and we extracted the Pauli transfer matrix with the process fidelity of 0.65. As an application to a real algorithm, we experimentally demonstrated the Deutch-Jozsa algorithm using the cNOT gate.

We observe errors in the cNOT gate from the fidelity measurement. We speculate that the infidelity can be attributed to 1) the decoherence during the long operation time of the cNOT gate, and 2) the leakage of the population into higher non-computational states mediated by the strong Stark tone. As a simple, conservative estimation of the decoherence induced infidelity, we consider our cNOT gate time of 1.47 $\mu$s (1.07 $\mu$s for the MAP gate plus 400 ns for Z-gate) compared to the worst $T_2^{*}$ of 5 $\mu$s in Q1. In this case, we would lose about 25\% ($\sim$ exp[-1.47/5]) of contrast of the Ramsey fringe after the total gate time of the cNOT gate.  Alternatively, since our Z-gate has quite high fidelity ($>$0.95 including SPAM), we may be able to include only the MAP gate time of 1.07 $\mu$s for major dephasing process, which gives about 20\% ($\sim$ exp[-1.07/5]) loss of contrast  after the MAP gate. These estimations roughly explain the majority of the infidelity we observed, but not all. We also need to note that the strong Stark tone we applied for the MAP gate also induces loss of Ramsey contrast, compared to a weak Stark tone drive. We infer this Stark-tone induced infidelity is due to the leakage of the population to the higher non-computational states. (In addition to shifting levels, the Stark tone drives transition from $
|01\rangle$ and $|11\rangle$ states to $|02\rangle$ and $|12\rangle$ states non-resonantly.) One obvious strategy to overcome both of these problems is to increase the coherence time of the qubits with the same setup, which is always possible. An alternative strategy is to enhance the vacuum Rabi coupling $g$, hence to increase the interaction between two qubits and the splitting between $|12\rangle$ and $|03\rangle$ states. Then a Stark tone with a lower power can lead to a reasonably shorter MAP gate time $t_g$ compared to the available coherence time without such a large leakage to higher states. 

\section*{Acknowledgements}

This research was supported by the R\&D convergence Program of NST (National Research Council of Science and Technology) of Republic of Korea (Grant No. CAP-15-08-KRISS).
\\
\\
\\

\begin{figure}[h]
\centering
\includegraphics[width=16cm]{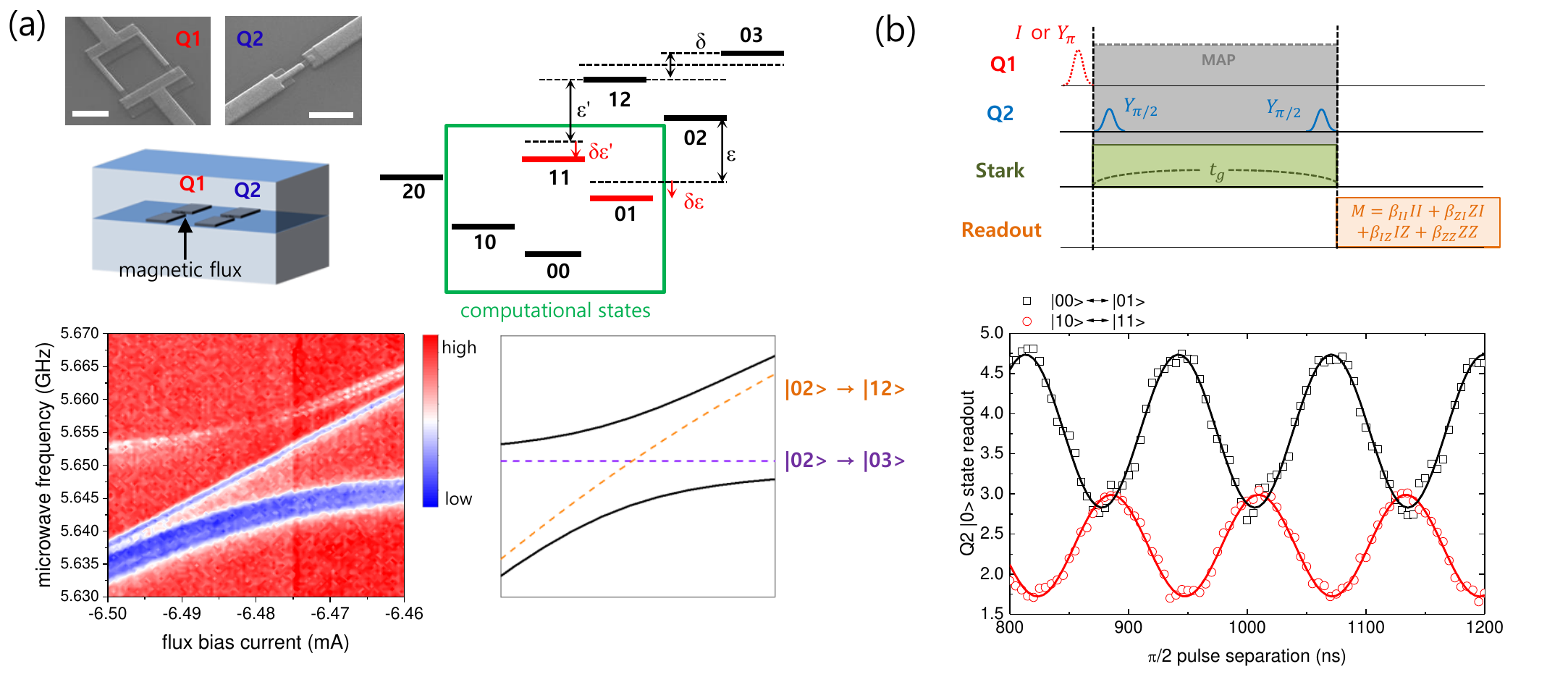}%
\caption{(a) Top left: Scanning electron microscope image of two qubits (scale bar is 1$\mu$m) with schematic diagram of the three dimensional circuit QED system configuration. Top right: Energy diagram of the two-qubit system. Four states inside the green solid box compose the computational states. The states $|12\rangle$ and $|03\rangle$ are tuned $\textit{in situ}$ by adjusting the magnetic flux bias through Q1. Bottom Left: The experimental result of three-tone microwave spectroscopy. The color plot shows the $|02\rangle$ population. $|12\rangle$ and $|03\rangle$ states are aligned at the current bias value of -6.48mA. Bottom right: Schematic representation of transitions induced by three-tone microwave spectroscopy. Orange (purple) dashed line corresponds to the transition from $|02\rangle$ to $|12\rangle$ (from $|02\rangle$ to $|03\rangle$). Due to the avoided crossing of two transitions, solid black lines are observed in the experiment. The straight line in between appears due to the transition from $|00\rangle$ to $|10\rangle$ and from $|01\rangle$ to $|11\rangle$, to which the reduction of the population at $|02\rangle$ state is attributed. (b) Top : Illustration of the pulse sequence of the MAP gate. With the $\pi$ pulse for Q1 on (off), the Ramsey fringe between $|00\rangle$ and $|01\rangle$ ($|10\rangle$ and $|11\rangle$) is induced by a Stark tone. Bottom : Experimental result of the Stark tone-induced Ramsey fringe. Here the $|00\rangle$ state population is measured for $|00\rangle$-$|01\rangle$ Ramsey, and the $|10\rangle$ state population for $|10\rangle$-$|11\rangle$ Ramsey. The oscillation period is determined by the amount of the shift $\delta \epsilon$ for the state $|01\rangle$ ($\delta \epsilon'$ for $|11\rangle$) as shown in (a). At $t_g$ = 1070 ns, two traces are completely out-of-phase.}
\end{figure}

\begin{figure}[ht]
\centering
\includegraphics[width=14cm]{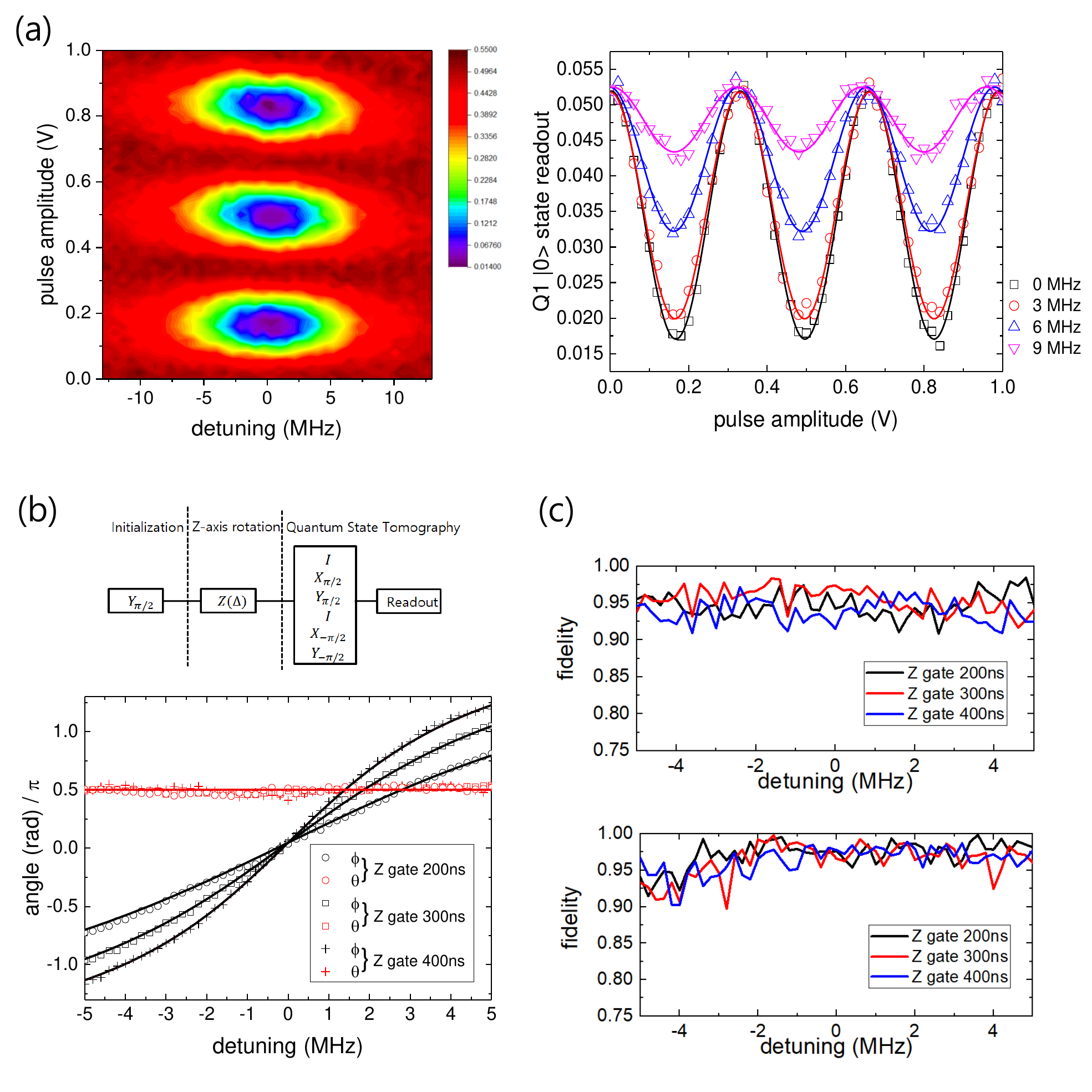}%
\caption{(a) Left: Detuning map of Rabi oscillation of Q1 driven by a microwave tone with hyperbolic secant pulse. Right: Rabi oscillations (taken from the left) with four different values of detuning $\Delta$, 0 MHz, 3 MHz, 6 MHz, and 9 MHz. Here the pulse length was fixed to be 400 ns. (b) Demonstration of the Z-axis phase gate on Q1 based on hyperbolic secant pulse. The schematic sequence to perform a single-qubit QST after applying a Z-axis phase gate is illustrated above the plot. The results of the polar angle $\theta$ (red circle, square, and triangle) and the azimuthal angle $\phi$ (black circle, square, and triangle) for three different pulse lengths 200 ns, 300 ns, and 400 ns of $2\pi$ cyclic rotation in the range of detuning -5 MHz $\leq \Delta \leq$ 5 MHz are shown in the figure. (c) Estimated state fidelity for each quantum state constructed in (b) by performing single-qubit QST for Q1 (upper panel) and Q2 (lower panel).}
\end{figure}

\begin{figure}[h]
\centering
\includegraphics[width=15cm]{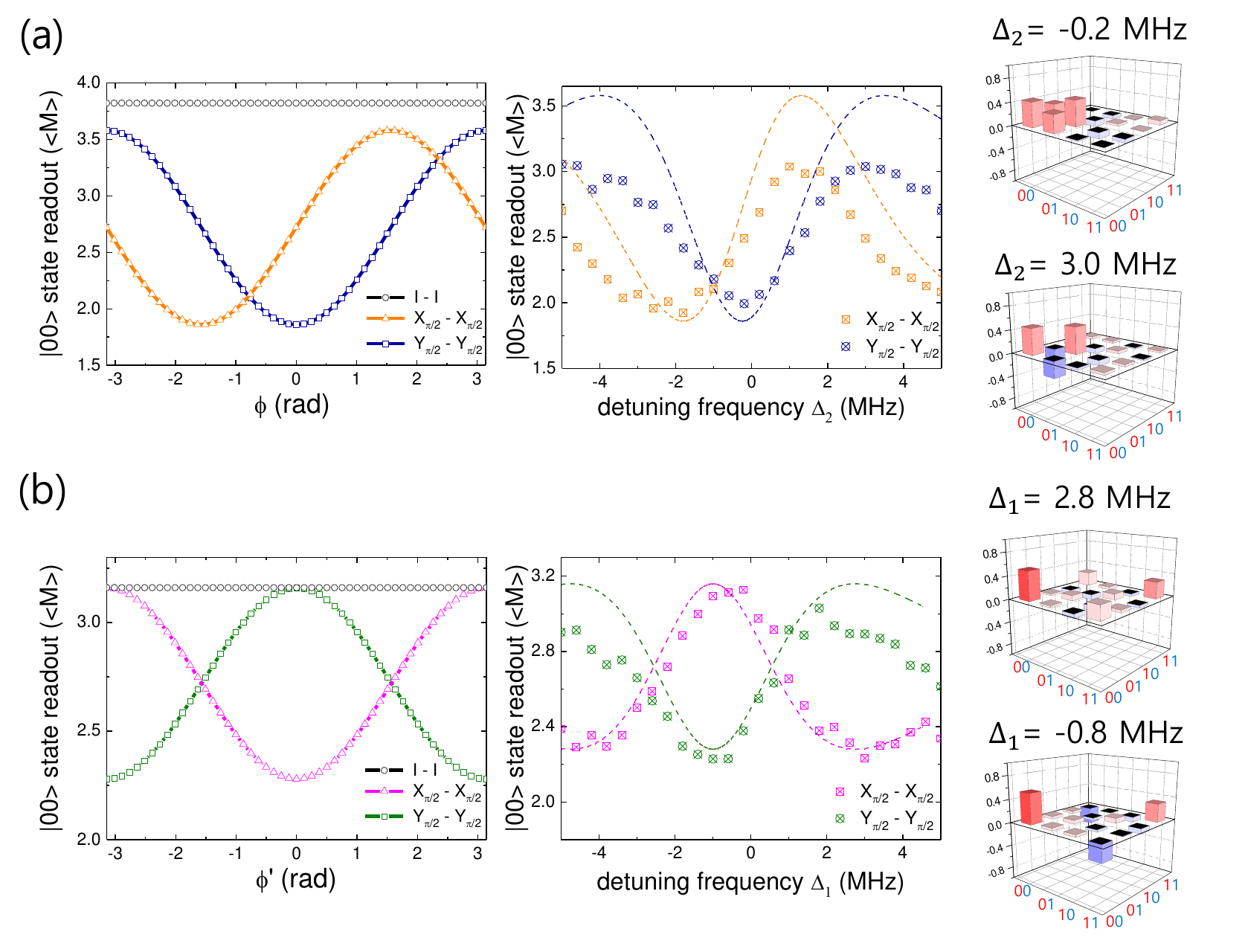}%
\caption{
(a) Left : Theoretical estimation of applying protocols of three different pre-pulses,  $I \otimes I$ (black open circle), $X_{\pi/2} \otimes X_{\pi/2}$ (orange open triangle) and $Y_{\pi/2} \otimes Y_{\pi/2}$ (navy open square) on the state $\frac{1}{\sqrt{2}} (|00\rangle+e^{i \phi}|01\rangle)$.  
Middle : Experimental result of applying protocols of pre-pulses $X_{\pi/2} \otimes X_{\pi/2}$ (orange cross square) and $Y_{\pi/2} \otimes Y_{\pi/2}$ (navy cross circle). Dotted lines are borrowed from the left. Right : The reconstructed density matrix by two-qubt QST. $\Delta_2$ = -0.2 MHz (top) and 3.0 MHz (bottom). (See the main text for the detailed explanation)
(b) Left : Theoretical estimation of applying protocols of three different pre-pulses, $I \otimes I$(black open circle), $X_{\pi/2} \otimes X_{\pi/2}$ (pink open triangle) and $Y_{\pi/2} \otimes Y_{\pi/2}$ (green open square) on the state $\frac{1}{\sqrt{2}} (|00\rangle+e^{i \phi'}|11\rangle)$.  Middle : Experimental result of applying protocols of pre-pulses $X_{\pi/2} \otimes X_{\pi/2}$ (pink cross square) and $Y_{\pi/2} \otimes Y_{\pi/2}$ (green cross circle). Dotted lines are borrowed from the left. Right : The reconstructed density matrix by two-qubt QST. $\Delta_1$ = 2.8 MHz for the top and -0.8 MHz for the bottom with $\Delta_2$ is fixed at -0.2 MHz as determined above. (See the main text for the detail)}
\end{figure}

\begin{figure}[ht]
\centering
\includegraphics[width=13cm]{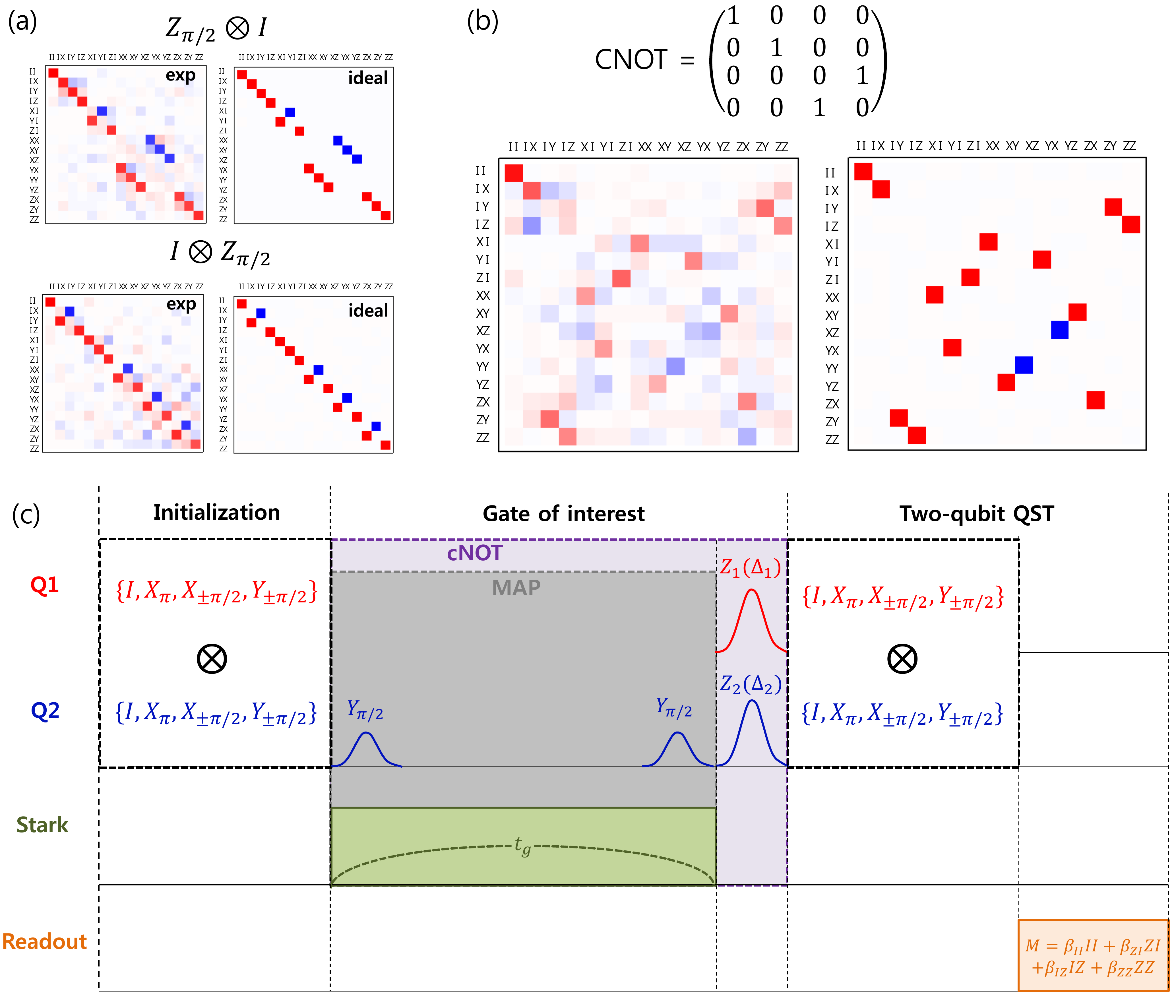}%
\caption{(a) Pauli transfer matrix $\textit{\textbf{R}}$  of the single-qubit gate $Z_{\pi/2} \otimes I$ and $I \otimes Z_{\pi/2}$ by two-qubit QPT. For each gate, ideal $\textit{\textbf{R}}$ is represented along with the experimental $\textit{\textbf{R}}$ for comparison. (b) Pauli transfer matrix $\textit{\textbf{R}}$ of the cNOT gate reconstructed from  experimental data (left) along with the ideal matrix (right). (c) Schematic diagram of the pulse sequence for two-qubit QPT in case of the cNOT gate. }
\end{figure}

\begin{figure}[ht]
\centering
\includegraphics[width=13cm]{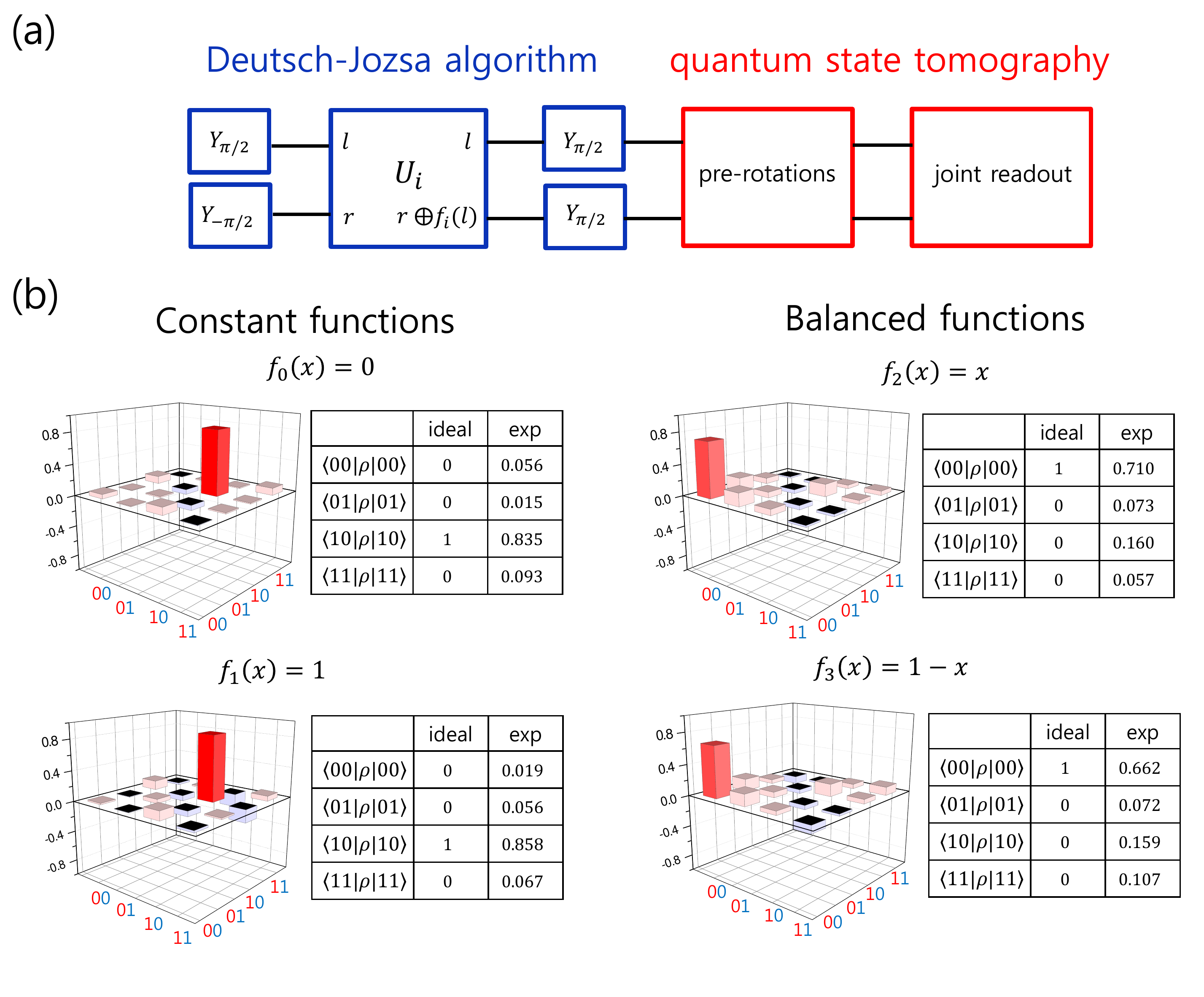}%
\caption{(a) Gate sequence for the demonstration of the Deutsch-Jozsa algorithm. (b) The real part of the density matrix of the output state for four encoding unitaries, $U_0 = I \otimes I$, $U_1 = I \otimes X_{\pi}$
$U_2 =\rm{cNOT}$, and $U_3 = (Y_{\pi} \otimes I){\rm{cNOT}}(Y_{-\pi} \otimes I)$. }
\end{figure}


\begin{thebibliography}{5}
\bibitem{barends} Barends, R. {\it{et al.}} Superconducting quantum circuits at the surface code threshold for fault tolerance. {\it{Nature}} {\bf{508}}, 500-503 (2014).
\bibitem{sheldon1} Sheldon, S., Bishop, L. S., Magesan, E., Filipp, S., Chow, J. M. \& Gambetta, J. M. Characterizing errors on qubit operations via iterative randomized benchmarking. {\it{Phys. Rev. A}} {\bf{93}}, 012301 (2016).
\bibitem{chen} Chen, Z. {\it{et al.}} Measuring and Suppressing Quantum State Leakage in a Superconducting Qubit. {\it{Phys. Rev. Lett.}} {\bf{116}}, 020501 (2016).
\bibitem{sheldon2} Sheldon, S., Magesan, E., Chow, J. M. \& Gambetta J. M. Procedure for systematically tuning up cross-talk in the cross-resonance gate. {\it{Phys. Rev. A}} {\bf{93}}, 060302(R) (2016).
\bibitem{corcoles} C\'{o}rcoles, A. D. {\it{et al.}} Process verification of two-qubit quantum gates by randomized benchmarking. {\it{Phys. Rev. A}} {\bf{87}}, 030301 (2013).
\bibitem{nori1} Gu, X., Kockum, A. F., Miranowicz, A., Liu, Y. X., \& Nori, F.  Microwave photonics with superconducting quamtum circuits. {\it{Physics Reports}} {\bf{718-719}}, 1-102 (2017).
\bibitem{dic1} Dicarlo, L. {\it{et al.}} Demonstration of two-qubit algorithms with a superconducting quantum processor. {\it{Nature}} {\bf{460}}, 240 (2009).
\bibitem{yama} Yamamoto, T. {\it{et al.}} Quantum process tomography of two-qubit controlled-Z and controlled-NOT gates using superconducting phase qubits. {\it{Phys. Rev. B}} {\bf{82}}, 184515 (2010). 
\bibitem{mariantoni} Mariantoni, M. {\it{et al.}} Implementing the Quantum von Neumann Architecture with Superconducting Circuits. {\it{Science}} {\bf{334}}, 61 (2011).
\bibitem{strau} Strauch, F. W., Johnson, P. R., Dragt, A. J., Lobb, C. J., Anderson, J. R. \& Wellstood, F. C. Quantum Logic Gates for Coupled Superconducting Phase Qubits. {\it{Phys. Rev. Lett}} {\bf{91}}, 167005 (2003). 
\bibitem{corc} C\'{o}rcoles, A. D. {\it{et al.}} Process verification of two-qubit quantum gates by randomized benchmarking. {\it{Phys. Rev. A}} {\bf{87}}, 030301(R) (2010). 
\bibitem{parao} Paraoanu, G. S. Microwave-induced coupling of superconducting qubits {\it{Phys. Rev. B}} {\bf{74}}, 140504 (2006). 
\bibitem{rigetti} Rigetti, C. \& Devoret, M. Fully microwave-tunable universal gates in superconducting qubits with linear couplings and fixed transition frequencies. {\it{Phys. Rev. B}} {\bf{81}}, 134507 (2010). 
\bibitem{chow3} Chow, J. M. {\it{et al.}} Simple All-Microwave Entangling Gate for Fixed-Frequency Superconducting Qubits. {\it{Phys. Rev. Lett.}} {\bf{107}}, 080502 (2011). 
\bibitem{paik} Paik, H. {\it{et al.}} Observation of High Coherence in Josephson Junction Qubits Measured in a Three-Dimensional Circuit QED Architecture. {\it{Phys. Rev. Lett.}} {\bf{107}}, 240501  (2011).
\bibitem{paik2} Paik, H. {\it{et al.}} Experimental Demonstration of a Resonator-Induced Phase Gate in a Multiqubit Circuit-QED System. {\it{Phys. Rev. Lett.}} {\bf{117}}, 250502 (2016).
\bibitem{chow1} Chow, J. M., Gambetta, J. M., Cross, A. W., Merkel, S. T., Rigetti, C. \& Steffen, M. Microwave-activated conditional-phase gate for superconducting qubits. {\it{New J. Phys.}} {\bf{15}}, 115012 (2013).
\bibitem{poletto} Poletto. S. {\it{et al.}} Entanglement of Two Superconducting Qubits in a Waveguide Cavity via Monochromatic Two-Photon Excitation. {\it{Phys. 
Rev. Lett.}} {\bf{109}}, 240505 (2012).
\bibitem{ku} Ku, H. {\it{et al.}} Single qubit operations using microwave hyperbolic secant pulses. {\it{Phys. Rev. A}} {\bf{96}}, 042339 (2017).
\bibitem{economou} Economou, S. E. \& Barnes, E. Analytical approach to swift nonleaky entangling gates in superconducting qubits, {\it{Phys. Rev. B}} {\bf{91}}, 161405(R) (2015)
\bibitem{filipp} Filipp, S. {\it{et al.}} Two-Qubit State Tomography Using a Joint Dispersive Readout. {\it{Phys. Rev. Lett.}} {\bf{102}}, 200402 (2009).
\bibitem{chow4} Chow, J. M. {\it{et al.}} Detecting highly entangled states with a joint qubit readout. {\it{Phys. Rev. A}} {\bf{81}}, 062325 (2010).
\bibitem{chow2} Chow, J. M. {\it{et al.}} Universal Quantum Gate Set Approaching Fault-Tolerant Thresholds with Superconducting Qubits. {\it{Phys. Rev. Lett.}} {\bf{109}}, 060501 (2012).
\bibitem{dj} Deutsch, D \& Jozsa, R. Rapid solution of problems by quantum computation. {\it{Proc. R. Soc. Lond. A}} {\bf{439}}, 553 (1992)
\end{thebibliography}
\end{document}